\documentclass[11pt]{article}

\usepackage[utf8]{inputenc}
\usepackage[]{graphicx}
\usepackage[squaren]{SIunits}
\usepackage{color}

\author{Denis Bernard, 
Philippe Bruel, 
Mickael Frotin\footnote{Now at  GEPI, Observatoire de Paris, CNRS, Univ. Paris Diderot, Place Jules Janssen, 92190 Meudon, France},
Yannick Geerebaert, 
Berrie Giebels, 
\\
Philippe Gros, 
Deirdre Horan, 
Marc Louzir,
Frédéric Magniette,
\\
Patrick Poilleux, 
Igor Semeniouk, 
Shaobo Wang 
$^{a}$ 
\\
$^{a}$ {\em LLR, Ecole Polytechnique, CNRS/IN2P3, 91128 Palaiseau, France}
\\
~
\\
David Attié,
Pascal Baron,
David Baudin,
Denis Calvet,
\\
Paul Colas, 
Alain Delbart, 
Patrick Sizun,
Ryo Yonamine
$^{b}$ 
\\
$^{b}$ {\em IRFU, CEA Saclay, 91191 Gif-sur-Yvette, France}
\\
~
\\
Diego Götz
$^{b,c}$
\\
$^{c}$ {\em AIM, CEA/DSM-CNRS-Université Paris Diderot, IRFU/Service d'Astrophysique,  }
\\ {\em 
CEA Saclay, F-91191 Gif-sur-Yvette, France }
\\
~
\\
Sho~Amano,
Satoshi~Hashimoto,
Takuya Kotaka,
Yasuhito~Minamiyama,
\\
Akinori Takemoto,
Masashi~Yamaguchi,
Shuji~Miyamoto
$^{e}$ 
\\
$^{e}$ {\em LASTI, University of Hy\={o}go, 3-1-2 Koto, Kamigori-cho, Ako-gun, Hy\={o}go 678-1205, Japan}
\\
~
\\
Schin Dat\'e and Haruo Ohkuma
$^{f}$ 
\\
$^{f}$ {\em  JASRI, 1-1-1, Kouto, Sayo-cho, Sayo-gun, Hy\={o}go 679-5198 Japan.}
}

\title{HARPO: 1.7 - 74 MeV gamma-ray beam validation of a high angular resolution, high linear polarisation dilution, gas time projection chamber telescope and polarimeter}

\begin{document}

\maketitle

\begin{abstract}
A presentation at the SciNeGHE conference of the past achievements, of
the present activities and of the perspectives for the future of the
HARPO project, the development of a time projection chamber as a
high-performance gamma-ray telescope and linear polarimeter in the
$e^+e^-$ pair creation regime.
\end{abstract}

A number of groups are developing pair-conversion detector
technologies alternative to the tungsten-converter /
thin-sensitive-layer stacks of the COS-B / EGRET / Fermi-LAT series,
to improve the single-photon angular resolution.
Presently observers are almost blind in the 1-100 MeV energy range,
mainly due to the degradation of the angular resolution of $e^+e^-$ pair
telescopes at low energies: to a large extent, the sensitivity-gap
problem is an angular-resolution issue.
Also no $\gamma$-ray polarimeter in the pair creation regime, that is,
above 1\,MeV, had ever been flown to space.

\begin{figure} [ht]
\begin{center}
 \includegraphics[width=0.6\linewidth]{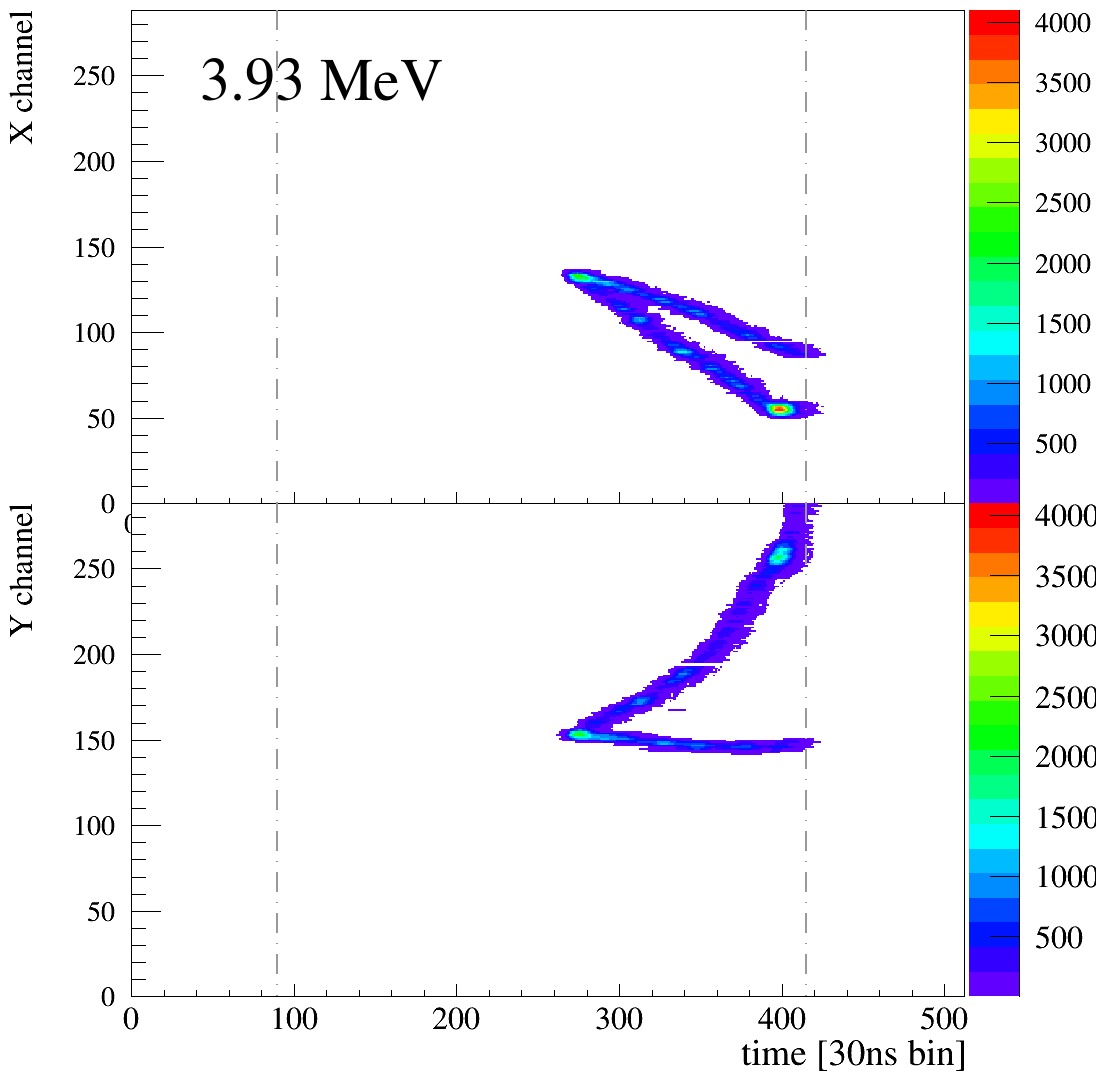}
 \caption{\label{fig:evt}
The two projections $(x,t)$ and $(y,t)$ of the signal recorded upon
the conversion of a 4 MeV photon provided by the BL01 polarized
$\gamma$-ray beam line at NewSUBARU, in the $2.1\,\bbar$
argon-isobutane 95-5\% gas mixture of the HARPO TPC.
The left and right dashed vertical lines correspond to the time of the
trigger and to the arrival time of the ``ionisation'' electrons who
enjoyed the full drift length from the cathode to the anode,
respectively. }
\end{center}
\end{figure}

Gas detectors such as TPCs (time projection chambers) can enable an
improvement of up to one order of magnitude in the single-photon
angular resolution (0.5\degree @ 100 MeV) with respect to the
Fermi-LAT (5\degree @ 100 MeV), a factor of three better than what can
be expected for Silicon detectors (1.0-1.5\degree @ 100
MeV) \cite{Bernard:2012uf}.
With such a good angular resolution, and despite a lower sensitive
mass, a TPC can contribute to close the sensitivity gap at the level
of $10^{-6}\mega\electronvolt/(\centi\meter^2
\second)$ \cite{Bernard:2012uf}.
In addition, the single-track angular resolution is so good that the
linear polarisation fraction and angle of the incoming radiation can
be measured \cite{Bernard:2013jea}. 
In constrast with dense telescopes (Si, emulsions), gas detectors can
detect the conversion of very low energy photons
(Fig. \ref{fig:evt}),
which is crucial as the signal of conversion to pairs peaks at
5 -- 7\,MeV for most cosmic sources.

The HARPO prototype is a $30\,\centi\meter$ cubic TPC. It uses an
argon-isobutane gas mixture as its active target, that is a fast-gas
with a full-length drift time of $9\micro\second$ that ensures a low
background pile-up probability in flight operation, and is also a
cool-gas that provides diffusion coefficients almost as low as the
thermal limit \cite{Bernard:2014kwa}.

We have characterized the detector performance in the pressure range
$1 - 4\,\bbar$ in the 1.7-74 MeV fully-polarised or non-polarised
gamma-ray beam provided by the BL01 line at
NewSUBARU \cite{Delbart:ICRC:2015}.
A trigger adapted to the large flux of single-track background noise
coming from the beam line was instrumental to enabling a
high-efficiency data taking
\cite{Yannick:RT2016}.
The excellent value of the polarisation asymmetry dilution factor that
we measured paves the way to the opening of the polarimetry window in
the MeV-GeV energy range \cite{Gros:SPIE:2016}.

In telescopes in which $\gamma$-ray conversions take place in high-$Z$
converter plates, such as in the Fermi LAT, the angular resolution is
coarse enough that the accuracy of the $\gamma \to e^+ e^-$ event
generator used in the simulation is not an issue.
But for the high angular resolution telescope that is considered here, the
kinematic contribution of the unmeasured ion recoil dominates the
single-photon angular resolution and the use of an exact generator,
such as the one that we developed \cite{Bernard:2013jea}, is
mandatory \cite{Gros:2016zst,PhilippeSciNeGHE}.
Also we have discovered that none of the otherwise existing 
generators simulate the conversion of linear polarization accurately
\cite{Gros:2016zst}.
We are presently editing our generator to make it available to the community.

With this tool, we found that the definition of ``the'' azimuthal
angle of the gamma conversion that was used since the end of the last
century in linear polarimetry studies, provides a measure of the
polarisation asymmetry that underestimates significantly its value,
which implies that the precision of the measurement of the
polarisation fraction of a source is then degraded (increased).
The maximal value of the asymmetry, and that is found to be compatible
with the low-energy and high-energy asymptotes, was obtained with the
azimuthal angle of the bisectrix of the electron and of the positron
directions \cite{Gros:2016dmp,PhilippeSciNeGHE}.

\begin{figure} [ht]
\begin{center}
 \includegraphics[width=0.6\linewidth, trim=30cm 11cm 38cm 15cm, clip]{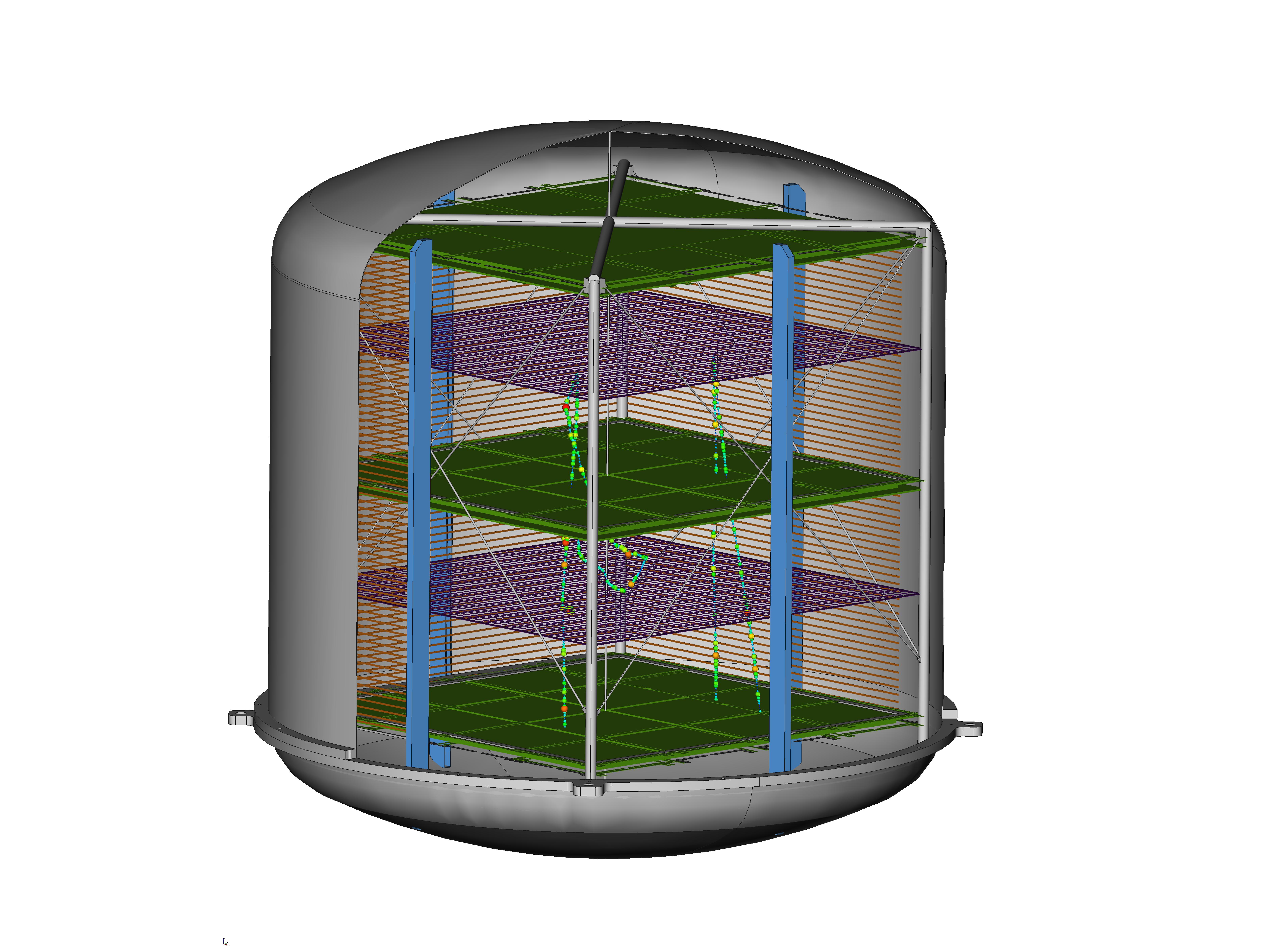}
 \caption{\label{fig:ST3G}
  An artist view of the flight prototype with 2 simultaneous 11.8\,MeV geant4 photon conversions overlaid.
 }
\end{center}
\end{figure}

The electrical power budget is strictly limited onboard space missions and
therefore it was not possible to equip the TPC endplate with a
collection anode with a 2D segmentation (pads).
Instead we use a pair of orthogonal series of strips ($x$ and $y$), at
the cost an ambiguity in the pairing of the two $x$ tracks and of the
two $y$ tracks.
The violent fluctuation of the ionisation energy deposition along the
tracks makes the signal time spectrum a  marker of each track,
that is collected almost in the same way by the $x$ strips and  by the $y$
strips, enabling an powerful track
matching \cite{Bernard:2012jy,Gros:SPIE:2016}.
In the special situation of two tracks almost overlapping in a given view
(say, $x$), obviously no matching is needed.
These events make peaks at azimutal angles $0$ and $ \pi/2$ which may be
frightening at first glance, but the azimutal information carried by
the pair is barely degraded as the dilution of the polarisation
asymmetry only decreases as $ e^{-2 \sigma_{\phi}^2}$, where $\sigma_{\phi}$ is the experimental resolution of the measurement of the azimuthal angle $\phi$ 
\cite{Bernard:2013jea}: a moments method enables the optimal extraction of
the available  azimuthal information \cite{Bernard:2013jea,Gros:2016dmp}.

The HARPO project has also implied several hardware developments; as
the gas quality of the EGRET tracker/converter degraded rapidly to the
extent that they had to change gas every year -- which is not
surprising for a spark chamber, most likely among the most efficient
way to perform chemistry under ionizing radiation -- we have monitored
the evolution of the gas quality of the HARPO detector when kept in a
sealed mode over several months: after a small oxygen component is
filtered out, the detector is functioning
nominally \cite{Frotin:2015mir}.

The ``ground'', HARPO, phase of the project is not far from
completion. We are now designing a ``flight'' prototype named ST3G
(stègue) ``Self-Triggered TPC for $\gamma$-ray telescope''
(Figure \ref{fig:ST3G})
for which
the trigger will be built using the real-time information provided by the
readout chips.
We have designed ASTRE ``Asic with SCA and Trigger for detector
Readout Electronics'' \cite{ASTRE} a modified version of the AGET
readout chip \cite{AGET}, that is also resistant to
ionizing radiation: a first batch of chips has been produced and
their tests are ongoing.
We aim at characterizing the behaviour of that flight trigger system
with a stratospheric balloon flight. People having interest to the
project or considering to join  are invited to attend the
``TPC for MeV Astrophysics'' workshop at Ecole Polytechnique, 12-14
April 2017.

The HARPO data taking on beam was performed by using NewSUBARU-GACKO
(Gamma Collaboration Hutch of Konan University).
It is a pleasure to acknowledge the support of the French National
Research Agency (ANR-13-BS05-0002).


\begin{thebibliography}{0}

\bibitem{Bernard:2012uf} 
 D.~Bernard,
 ``TPC in gamma-ray astronomy above pair-creation threshold,''
 Nucl.\ Instrum.\ Meth.\ A {\bf 701}, 225 (2013)
 Erratum: [Nucl.\ Instrum.\ Meth.\ A {\bf 713}, 76 (2013)]
 [arXiv:1211.1534 [astro-ph.IM]].

\bibitem{Bernard:2013jea} 
 D.~Bernard,
 ``Polarimetry of cosmic gamma-ray sources above $e^+e^-$ pair creation threshold,''
 Nucl.\ Instrum.\ Meth.\ A {\bf 729}, 765 (2013)
 [arXiv:1307.3892 [astro-ph.IM]].

\bibitem{Bernard:2014kwa} 
 D.~Bernard {\it et al.},
 ``HARPO: a TPC as a gamma-ray telescope and polarimeter,''
 Proc.\ SPIE Int.\ Soc.\ Opt.\ Eng.\ {\bf 9144}, 91441M (2014)
 [arXiv:1406.4830 [astro-ph.IM]].
 
\bibitem{Delbart:ICRC:2015}
A. Delbart {\it et al.},
 "HARPO, TPC as a gamma telescope and polarimeter: First measurement in a polarised photon beam between 1.7 and 74 MeV",
PoS(ICRC2015)1016.

\bibitem{Yannick:RT2016}
 Y. Geerebaert {\it et al.},
 "Electronics for HARPO, design, development and validation of electronics for a high performance polarized $\gamma$-ray detector", proceedings Real Time Conference (RT), 2016 IEEE-NPSS.
 
\bibitem{Gros:SPIE:2016}
P.~Gros {\it et al.},
``Measurement of polarisation asymmetry for gamma rays between 1.7 to 74 MeV with the HARPO TPC'',
SPIE2016, 9905-95.
 
\bibitem{Gros:2016zst} 
 P.~Gros and D.~Bernard,
 ``gamma-ray telescopes using conversions to electron-positron pairs: event generators, angular resolution and polarimetry,''
 Astropart.\ Phys.\  {\bf 88}(2017) 60,
 [arXiv:1612.06239 [astro-ph.IM]].
 
\bibitem{Gros:2016dmp}
 P.~Gros and D.~Bernard,
 ``$\gamma$-ray polarimetry with conversions to $e^+e^-$ pairs: polarization asymmetry and the way to measure it,''
 Astropart.\ Phys.\ {\bf 88} (2017) 30,
 [arXiv:1611.05179 [astro-ph.IM]].

\bibitem{PhilippeSciNeGHE}
 P.~Gros, contribution to this SciNeGHE conference.

\bibitem{Bernard:2012jy} 
  D.~Bernard,
 ``HARPO-A gaseous TPC for high angular resolution $\gamma$-ray astronomy and polarimetry from the MeV to the GeV,''
  12th Pisa Meeting on Advanced Detectors,
  Nucl.\ Instrum.\ Meth.\ A {\bf 718}, 395 (2013)
  [arXiv:1210.4399 [astro-ph.IM]].
  
\bibitem{Frotin:2015mir} 
 M.~Frotin {\it et al.},
 ``Sealed operation, and circulation and purification of gas in the HARPO TPC,''
 Proceedings MPGD2015, EPJ Web of Conferences,
 arXiv:1512.03248 [physics.ins-det].

\bibitem{ASTRE}
 O. Gevin,
 ``Développements récents à l'IRFU'',
 Journées VLSI-PCB-FPGA 2016, Strasbourg. 31 mai-8 juin 2016. 

\bibitem{AGET} 
S. Anvar, P. Baron {\it et al.},
{\it AGET, the GET front-end ASIC, for the readout of the Time Projection Chambers used in nuclear physic experiments},
Nuclear Science Symposium and Medical Imaging Conference (NSS/MIC), 2011 IEEE
745 - 749.
 
\end{thebibliography}
\end{document}